\documentclass{emulateapj}
\usepackage{apjfonts}

\def\be{\begin{equation}}
\def\ee{\end{equation}}

\newcommand{\msun}{{\rm M}_{\sun}}

\catcode`\@=11 
\def\@versim#1#2{\vcenter{\offinterlineskip
        \ialign{$\m@th#1\hfil##\hfil$\crcr#2\crcr\sim\crcr } }}

\shorttitle{****}
\shortauthors{****}
\begin{document}

\title{Global Compton heating and cooling in hot accretion flows}

\author{Feng Yuan\altaffilmark{1, 2}, Fuguo Xie\altaffilmark{1, 3},
Jeremiah P. Ostriker\altaffilmark{4}}
\altaffiltext{1}{Shanghai Astronomical Observatory, Chinese Academy of Sciences,
80 Nandan Road, Shanghai 200030, China; fyuan@shao.ac.cn}
\altaffiltext{2}{Joint Institute for Galaxy and Cosmology (JOINGC) of
SHAO and USTC}
\altaffiltext{3}{Graduate School of Chinese Academy of Sciences,
Beijing 100039, China}
\altaffiltext{4}{Princeton University Observatory, Princeton, USA}

\begin{abstract}

The hot accretion flow (such as advection-dominated accretion flow)
is usually optically thin in the radial direction,
therefore the photons produced at one radius
can travel for a long distance without being absorbed. These photons
thus can heat or cool electrons at other radii via Compton scattering.
This effect has been ignored in most previous works on hot accretion flows
and is the focus of this paper. If the mass accretion rate is described by
$\dot{M}=\dot{M}_0(r/r_{\rm out})^{0.3}$ and $r_{\rm out}=10^4r_{\rm s}$,
we find that the Compton scattering will play a cooling and heating role
at $r\la 5\times 10^3 r_{\rm s}$ and $r\ga 5\times 10^3 r_{\rm s}$, 
respectively. Specifically, when $\dot{M}_0>0.1L_{\rm Edd}/c^2$, the 
Compton cooling rate is larger than the local viscous heating rate at 
certain radius; therefore the cooling effect is important. 
When $\dot{M}_0>2L_{\rm Edd}/c^2$,
the heating effect at $r_{\rm out}$ is important. We can obtain
the self-consistent steady solution with the global Compton effect included
only if $\dot{M}_0\la L_{\rm Edd}/c^2$ for $r_{\rm out}=50r_{\rm s}$,
which corresponds to $L\la 0.02L_{\rm Edd}$. Above this rate the Compton
cooling is so strong at the inner region that hot solutions can not exist.
On the other hand, for $r_{\rm out}= 10^5r_{\rm s}$, we can only
get the self-consistent solution when $\dot{M}_0\la L_{\rm Edd}/c^2$ and 
$L<0.01L_{\rm Edd}$. The
value of this critical accretion rate is anti-correlated with the value of
$r_{\rm out}$. Above this accretion rate the equilibrium temperature of
electrons at $r_{\rm out}$ is higher than the virial temperature
as a result of strong Compton heating, so the accretion is suppressed.
In this case the activity of the black hole will likely ``oscillate''
between an active and an inactive phases, with the oscillation
timescale being the radiative timescale of the gas at $r_{\rm out}$.

\end{abstract}


\section{Introduction}

Compton scattering between photons and electrons
is an important process in astrophysics.
If the photons are not produced at the same place where the electrons
are located, we call it ``global Compton scattering''.
Momentum and energy of photons and electrons can be exchanged in this process
and these two aspects often play an
important role in determining the dynamics of the gas flow.
On the galactic scale, this so-called radiative
feedback mechanism now is believed to be
crucial for understanding AGN feedback on galaxy formation and evolution
(e.g., Ciotti \& Ostriker 2001, 2007;
Murray et al. 2005; Hopkins et al. 2005). On a smaller scale, the
effect of Compton scattering on the dynamics of gas flows surrounding
a strongly radiating quasar has been investigated and outflow is
produced as a consequence of Compton heating (e.g., Proga, Ostriker \&
Kurosawa 2008). Following the earlier work of
Krolik et al. (1981), Mathews \& Ferland (1987) considered the Compton heating
effect for the broad-line region of quasar.
This effect is also important for
the standard thin disk, if the disk is warped or 
irradiated by a source above the disk plane (e.g., Shakura \& Sunyaev 1973;
Begelman, McKee \& Shields 1983; Dubus et al. 1999). If the accretion flow is
geometrically thick and optically thin,
the photons can travel a large distance without being absorbed,
therefore the global Compton scattering effect is in principle important.
This is the case for spherical accretion and hot accretion flows. The latter
includes the advection-dominated accretion flow (ADAF; Narayan \& Yi 1994;
1995) and luminous hot accretion flow (LHAF; Yuan 2001; 2003),
two types of hot accretion flows corresponding to low and high accretion
rates, respectively.

For spherical accretion, the interaction of momentum between photons and
electrons sets up a largest possible luminosity the accretion flow
could reach, namely the Eddington luminosity $L_{\rm Edd}$
(but this limit does not apply when the accretion flow has
a non-zero angular momentum; see, e.g., Ohsuga \& Mineshige 2007).
The effect of the energy interaction between photons and electrons
in a spherical accretion flow has been investigated by Ostriker et al.
(1976). It was found that when the luminosity is
larger than a certain value the outward energetic
photons could heat gas flow significantly so that the local sound speed
is larger than the escape speed, or, in other words, the temperature
is higher than the virial temperature, thus the accretion is suppressed;
and this effect due to energy input from the outgoing radiation field
occurs for much lower luminosities than the momentum (Eddington) limit.

In almost all of the previous work on the dynamics of hot accretion
flows, only the ``local''  Compton scattering effect has been
considered while the global Compton effect has been neglected.
Here ``local'' means that photons are
produced at the same region with where the electrons locate. This
local Compton scattering serves as the main cooling mechanism of
electrons (Compton cooling) and the main mechanism of producing
X-ray emission (thermal Comptonization). The only works to our
knowledge considering the global Compton scattering are Esin (1997)
and Park \& Ostriker (1999; 2001; 2007). Esin (1997) deals with a
one-dimensional ADAF and finds that the global Compton
heating/cooling is not important and can be neglected. Park \&
Ostriker (2001; 2007) deal with a two-dimensional flow and focus
on the possible production of outflow in the polar region because of
strong Compton heating there. Their conclusion is that Compton
heating effect is important in many cases, which is quite different
from Esin's result. They do not attempt to obtain the
self-consistent solutions.

All their work are based on the self-similar solution of ADAF (Narayan \&
Yi 1994; 1995). The discrepancy between Esin and Park \& Ostriker
is likely due to the additional but different assumptions adopted.
While the self-similar approximation is quite
successful in catching the main spirit of an ADAF, it is not a good
approximation when we want to calculate the radiation since
order of magnitude error could be produced. This is because the self-similar
approximation breaks down
at the inner region of the ADAF where most of the radiation comes from.
When we consider the effect of global Compton heating/cooling,
obviously it is crucial to calculate the exact spectrum from the
exact global solution of the accretion flow. This is the main motivation
of the present paper. In addition, important theoretical progress on
ADAF solutions have been made since its discovery.
The two most important ones are the presence of outflow (e.g.,
Stone \& Pringle 2001) and significant direct
electron heating by turbulent dissipation (e.g., Quataert \& Gruzinov 1999),
which is much stronger than the heating by Coulomb collisions between ions
and electrons. Both of them are important in
determining the dynamics of ADAFs while they have not been properly
taken into account in the above works.
In the present paper we will focus on the one-dimensional case,
given that the uncertainty of our understanding to two-dimensional structure of
accretion flow is still large. Of course,
a full understanding to the two-dimensional case is obviously important
and should be a topic of future research.

\section{The importance of Compton heating or cooling in hot accretion flows}

\subsection{Calculation Method}

Consider a canonical hot accretion flow without taking into account the
global Compton heating or cooling. Outflow is taken into account by adopting
the following radius-dependent mass accretion rate (e.g.,
Blandford \& Begelman 1999):
\be
\dot{M}=-4\pi r H\rho v=\dot{M}_0\left(\frac{r}{r_{\rm out}}\right)^s,
\ee
where $\dot{M}_0$ is the mass accretion rate at the outer boundary
$r_{\rm out}$. The value of index $s$ describes the strength of the outflow
and we use $s=0.3$ from the detailed modeling of Sgr A*, the supermassive
black hole in our Galactic center (Yuan, Quataert \& Narayan 2003).
The energy equations for ions and electrons are
\be \rho v
\left(\frac{d \varepsilon_{\rm i}}{dr}- {p_{\rm i} \over \rho^2} \frac{d \rho}{dr}
\right) =(1-\delta)q^+-q_{\rm ie},
\ee
\be \rho v
\left(\frac{d \varepsilon_{\rm e}}{dr}- {p_{\rm e} \over \rho^2} \frac{d
\rho}{dr}\right) =\delta q^++q_{\rm ie}-q^-, \ee
where $\varepsilon_{\rm {e,i}}$
is the internal energy of electrons and ions per unit mass of the gas,
$q_{\rm ie}$ is the Coulomb energy exchange rate between electrons and ions,
$q^-$ is the electron cooling rate, including synchrotron and bremsstrahlung
emissions and their local Comptonization, $q^+$ is the net turbulent
heating rate, the value of $\delta$ describes the fraction of turbulent
heating which directly heats electrons and we use $\delta=0.5$ again
from the modeling of Sgr A* (see also Sharma et al. 2007).

We first get the exact global solution of the hot accretion flow so that
we know all the quantities such as density
and temperature as a function of radius. This requires us to solve
the set of equations describing the conservations of mass (eq. 1),
energy (eqs. 2 \& 3), and momentum (ref. Yuan, Quataert \& Narayan 2003).
The global solution should satisfy the outer boundary condition
at the outer boundary $r_{\rm out}$, the inner boundary
condition at the horizon, and a sonic point condition at the sonic point. To
calculate the rate of ``global'' Compton heating/cooling of electrons
at a given radius $r$, we need to know the spectrum received
at $r$ emitted by the whole
flow. This requires us to solve the radiative transfer equations along
the radial direction, which is complicated when scattering is important.
For simplicity, here we deal with the scattering in
a simple way and write the received spectrum at $r$ emitted by
the flow inside of $r$ as,
\be
F_{\nu}^{\rm in}(r)=\int^{r}_{r_{\rm s}}e^{-\tau}\frac{1}{4\pi r^2}
\frac{{\rm d}L_{\nu}(r')}{{\rm d}r'}{\rm d}r'
\ee
Here $\tau$ is the scattering
optical depth from $r'$ to $r$, $\tau=\int^r_{r'}\sigma_{\tiny T} n_e dr'$,
and $dL_{\nu}(r')$ is the emitted monochromatic luminosity from a shell at $r'$
with thickness $dr'$ and height $H(r')$. It includes synchrotron
and bremsstrahlung emissions and their local Comptonization.
We approximate the calculation of the unscattered part of
$dL_{\nu}(r')$ by solving the radiative transfer along the vertical direction
of ADAFs adopting a two-stream approximation
(Manmoto, Mineshige \& Kusunose 1997):
\be
dL_{\nu}^{\rm un}(r')=
\frac{4\pi^2}{\sqrt{3}}B_{\nu}[1-{\rm exp}(-2\sqrt{3}\tau_{\nu}^*)]r'dr'
\ee
Here $B_{\nu}$ denotes the Planck spectrum, $\tau_{\nu}^*\equiv
(\pi/2)^{1/2}\kappa_{\nu}(0)H(r')$ is the optical depth for absorption in the
vertical direction with $\kappa_{\nu}(0)$ being the absorption coefficient
on the equatorial plane. The free-free absorption and synchrotron
self-absorption are included in
this way. The strength of the magnetic field in the accretion flow is determined
by a parameter $\beta$ defined as the ratio of the gas pressure to the
magnetic pressure and we set $\beta=9$.
For the calculation of the Compton scattered part of $dL_{\nu}(r')$,
we use the approach of Coppi \& Blandford (1990; eq. 2.2).
The integration in eq. (4) begins from the black hole horizon $r_{\rm s}\equiv 2GM/c^2$.

The spectrum received at $r$ emitted by the
flow outside of $r$ is (Park \& Ostriker 2007),
\be
F_{\nu}^{\rm out}(r)=\int^{r_{\rm out}}_{r}\frac{e^{-\tau}}
{4\pi r' H(r')}\frac{r'}{r}
{\rm ln} \sqrt{\frac{r'+r}{r'-r}}\frac{dL_{\nu}(r')}{dr'}dr'
\ee
The total spectrum received at $r$ is the sum of $F_{\nu}^{\rm in}(r)$
and $F_{\nu}^{\rm out}(r)$.

\begin{figure} \epsscale{1.} \plotone{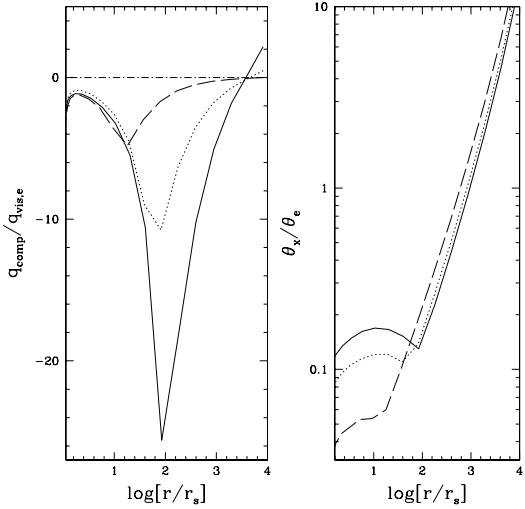} \vspace{.2in}
\caption{{\it Left panel}: The ratio of the Compton heating/cooling
rate to the turbulent heating rate of electrons in hot accretion
flows for $\dot{M}=\dot{M}_0(r/10^4r_s)^{0.3}$ with three accretion
rates $\dot{M}_0=0.1~\dot{M}_{\rm Edd}$ (dashed), $1~\dot{M}_{\rm
Edd}$ (dotted) and $2~\dot{M}_{\rm Edd}$ (solid). {\it Right panel}:
The ratio of the radiation temperature $\theta_{\rm x}$ (see text
for definition) and the electron temperature $\theta_e$.
When $\theta_{\rm x}$ is larger (smaller) than $\theta_e$,
Compton scattering plays a heating (cooling) role.}
\end{figure}

We assume that the electrons have a Maxwell distribution with
temperature $T_e$ ($\theta_e\equiv kT_e/m_ec^2$) and the energy of
the photon before scattering is $\epsilon \equiv h\nu/m_ec^2$. Since
the electrons in the hot accretion flow are relativistic at the
innermost region and the peak photon energy $\epsilon > 1$, to
calculate the average energy of a scattered photon, we use the
following exact form which is valid for any photon energy and
electrons temperature (Guilbert 1986): 
\begin{equation}
  \begin{array}{l}
<\epsilon_1>=\epsilon + \\
\frac{\sigma_T}{2K_2(1/\theta_e)\sigma}\int^{+\infty}_
{-\infty} \left(\theta_e+{\rm sinh}~\phi-\epsilon \right) G(\epsilon
e^{\phi})e^{2\phi} {\rm exp}\left(\frac{-{\rm cosh~\phi}}{\theta_e}
\right)d\phi, 
  \end{array}
\end{equation} with $G(\epsilon)\equiv
g_0(\epsilon)-g_1(\epsilon)$ and the cross-section for scattering
\be \sigma(\epsilon,\theta_e)=\frac{\sigma_T}{2K_2(1/\theta_e)}
\int^{+\infty}_{-\infty}g_0(\epsilon e^{\phi}) e^{2\phi}{\rm
exp}\left(\frac{-{\rm cosh}~\phi}{\theta_e}\right)d\phi. \ee Here
$K_2(x)$ is a modified Bessel function of second order and: \be
g_n(y) \equiv \frac{3}{8}\int^2_0\left( t(t-2)+1+ty+\frac{1}{1+ty}
\right)\frac{dt}{(1+ty)^{n+2}}. \ee In the Thompson limit, eqs. (7)
\& (8) are transformed into the familiar form of \be<\epsilon_1>
=\epsilon+\epsilon \frac{4kT_e-\epsilon~m_ec^2}{m_ec^2}, \ee and
\be\sigma(\epsilon,\theta_e)=\sigma_T.\ee

The number of scattering in a region of the accretion flow with unit
width in the radial direction and scattering optical depth
$\tau_{es}\equiv \sigma(\epsilon,\theta_e)n_e$ is \be N=\tau_{es},
\ee with $\theta_e$ and $n_e$ are the temperature and number density
of electrons in that region. The Compton heating (cooling) rate in
that region (with unit radial length) is then \be q_{\rm comp}=\int
N~[F_{\nu}^{\rm in}(r)+F_{\nu}^{\rm out}(r)]
\frac{\epsilon-<\epsilon_1>}{\epsilon} d\nu. \ee 
Note that in the above equation we actually use the moment of 
intensity ``$J$'' not ''$F$''. 
Following Park \& Ostriker (2007), we formally define a ``radiation 
temperature'' (or ``Compton temperature''; see also Levich \& Sunyaev 
1970; Krolik, McKee \& Tarter 1981): \be \theta_{\rm x}\equiv \frac{\int
[F_{\nu}^{\rm in}(r)+F_{\nu}^{\rm out}(r)]h\nu d\nu} {4m_ec^2\int
[F_{\nu}^{\rm in}(r)+F_{\nu}^{\rm out}(r)] d\nu}.\ee
Under this definition, whether the Compton scattering plays a
heating or cooling effect roughly depends on whether the electron
temperature $\theta_e$ is larger or smaller than $\theta_{\rm x}$.
In the Thompson limit, the Compton heating/cooling rate is exactly
proportional to $(\theta_{\rm x}-\theta_e)$.

It is important to note here that the radiation temperature
is obtained from the flux distribution by weighting with the factor
$h\nu$. It physically represents the equilibrium between Compton
heating and cooling. For a typical quasar spectrum, it corresponds
to $2\times 10^7~$K or several keV where the bulk of the
radiation is emitted in the UV or, in some cases, IR portions of the
spectrum (Mathews \& Ferland 1987; Sazonov, Ostriker \& Sunyaev 2004).
As we will see below (e.g., ref. Fig. 3 (c)), the radiation temperature of an
ADAF spectrum is much higher because of its different spectrum.

\subsection{Results}

\begin{figure*}
\epsscale{0.5} \plotone{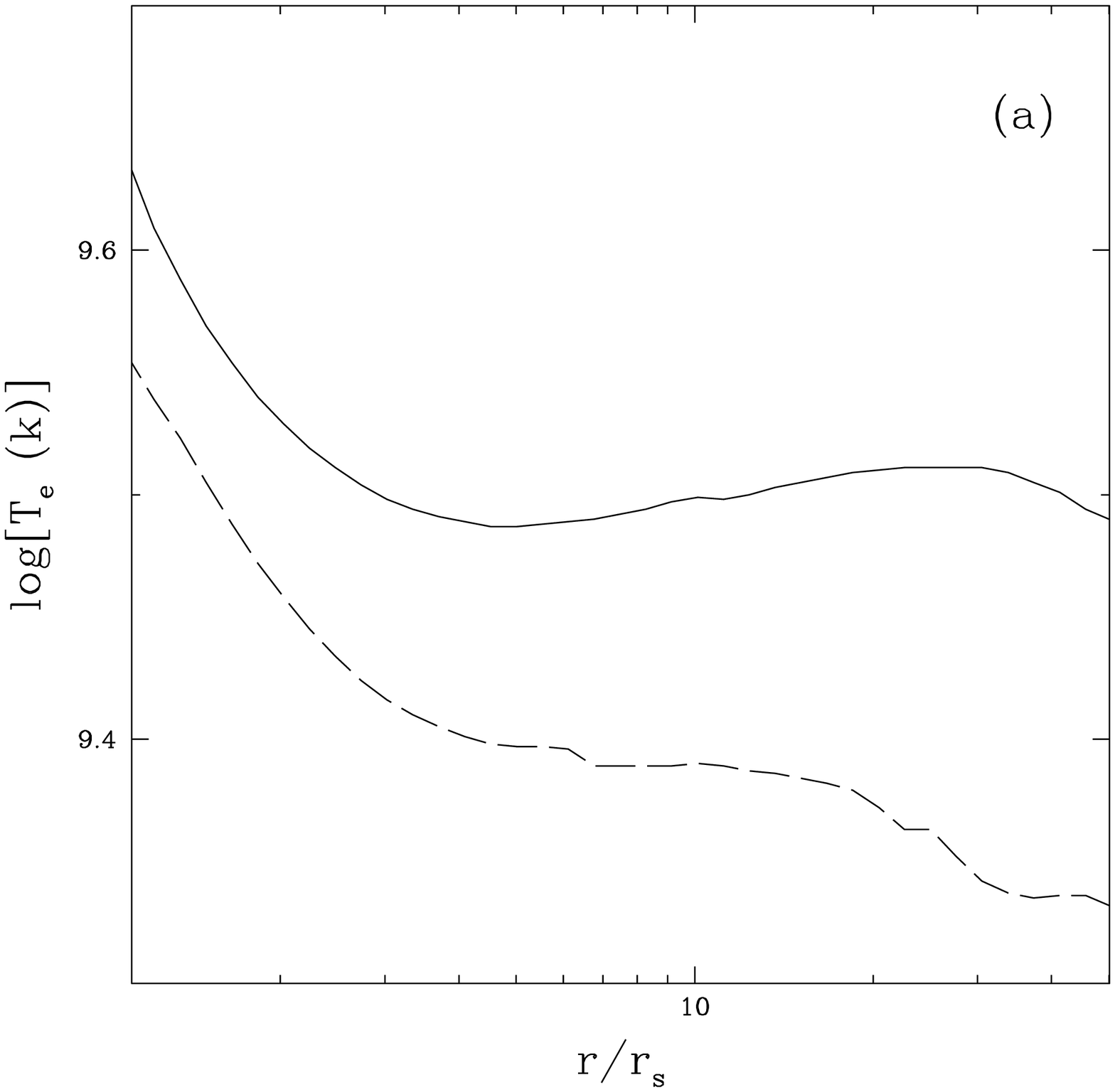} \\
\epsscale{0.45} \plotone{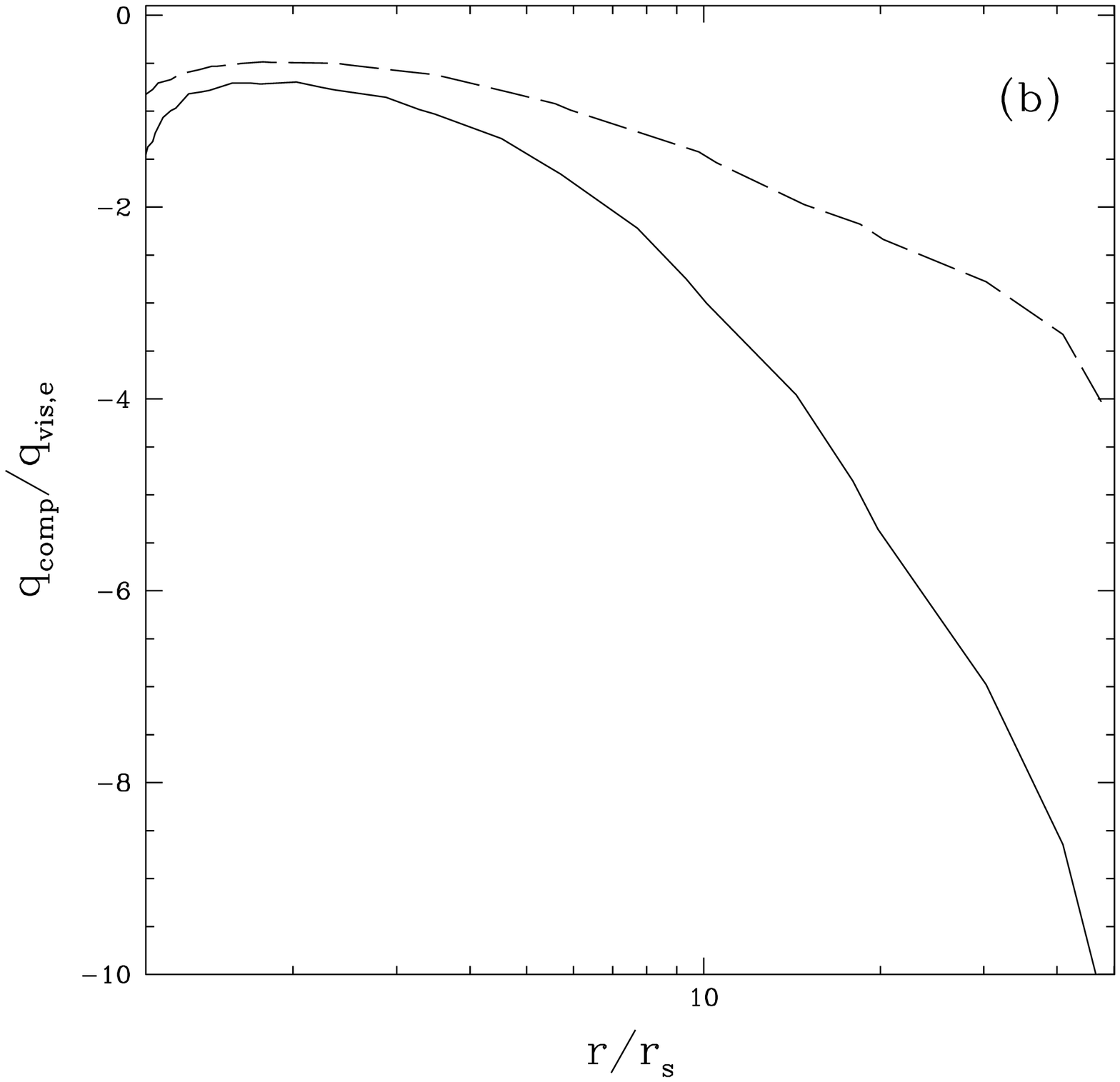} \epsscale{0.45} \plotone{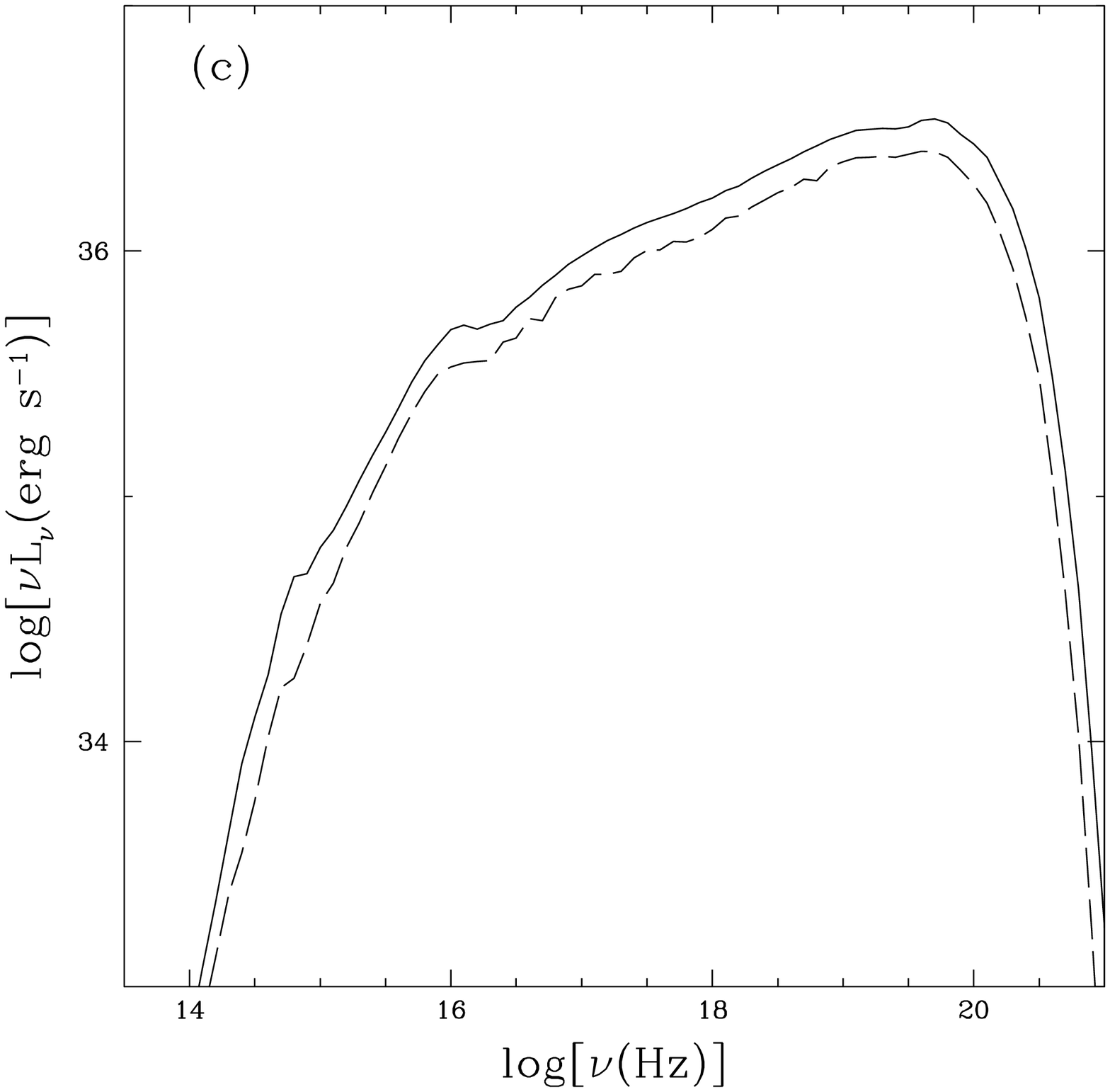}
\caption{The Compton effect for a stellar mass black hole with black hole
mass of $M=10~\msun$ and mass accretion rate
$\dot{M}=\left(\frac{r}{50r_s}\right)^{0.3}\dot{M}_{\rm Edd}$.
The solid and dashed lines are for the solutions before and after the global
Compton effect is taken into account. (a) The electron temperature profile.
(b) The ratio of the rate of Compton heating/cooling and the turbulent
heating of electrons. (c) The spectrum of the accretion flows.}
\end{figure*}

The dominant heating term of electrons in eq. (3) is $q_{\rm vis,e}
\equiv \delta q^+$. We compare the rate of global Compton heating/cooling
with $q_{\rm vis,e}$ and the results are shown in Fig. 1
for $\dot{M}=\dot{M}_0(r/10^4r_{\rm s})^{0.3}$ with
$\dot{M}_0=0.1, 1$, and $2\dot{M}_{\rm Edd}$ ($\dot{M}_{\rm Edd}
\equiv L_{\rm Edd}/c^2$). At large radii, $r\ga 5\times 10^3 r_{\rm s}$,
Compton scattering heats electrons; while at small radii,
$r\la 5\times 10^3r_{\rm s}$, it cools electrons. This is of course because the
radiation temperature $\theta_{\rm x}$ is lower (higher) than the electron
temperature $\theta_{\rm e}$ at the small (large) radii, as shown by
the right panel of Fig. 1.

\begin{figure*}
\epsscale{0.5} \plotone{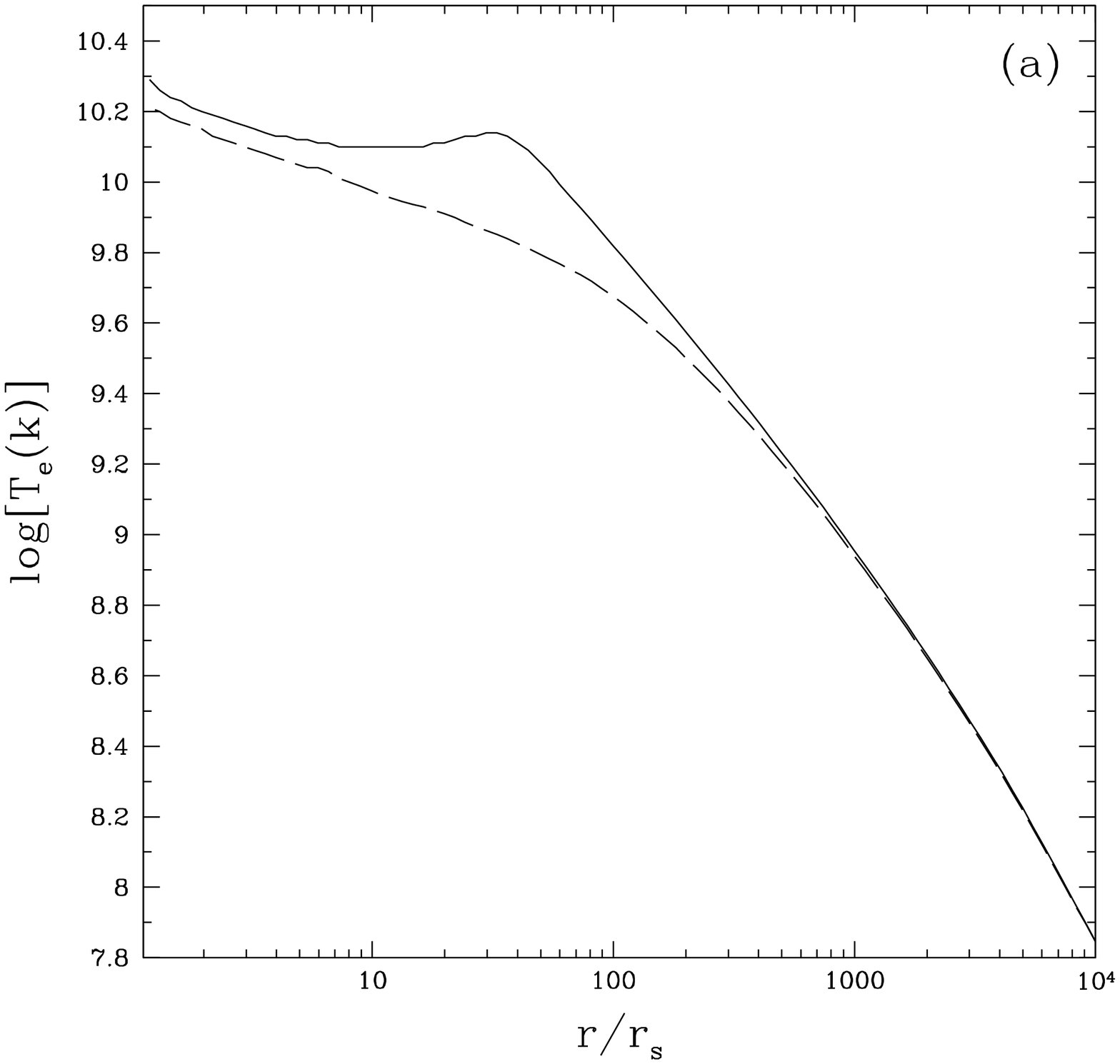} \\
\epsscale{0.45} \plotone{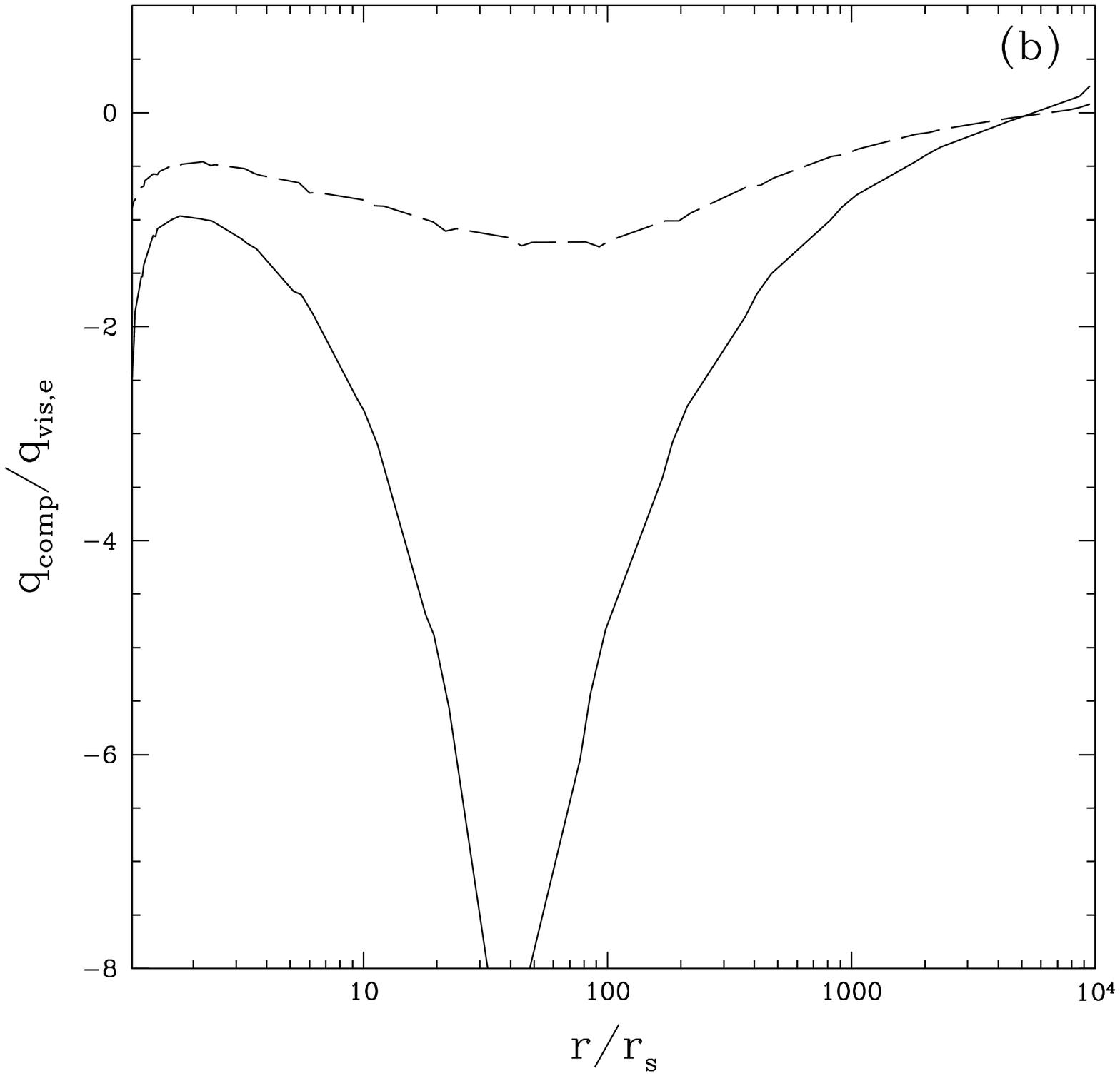} \epsscale{0.45} \plotone{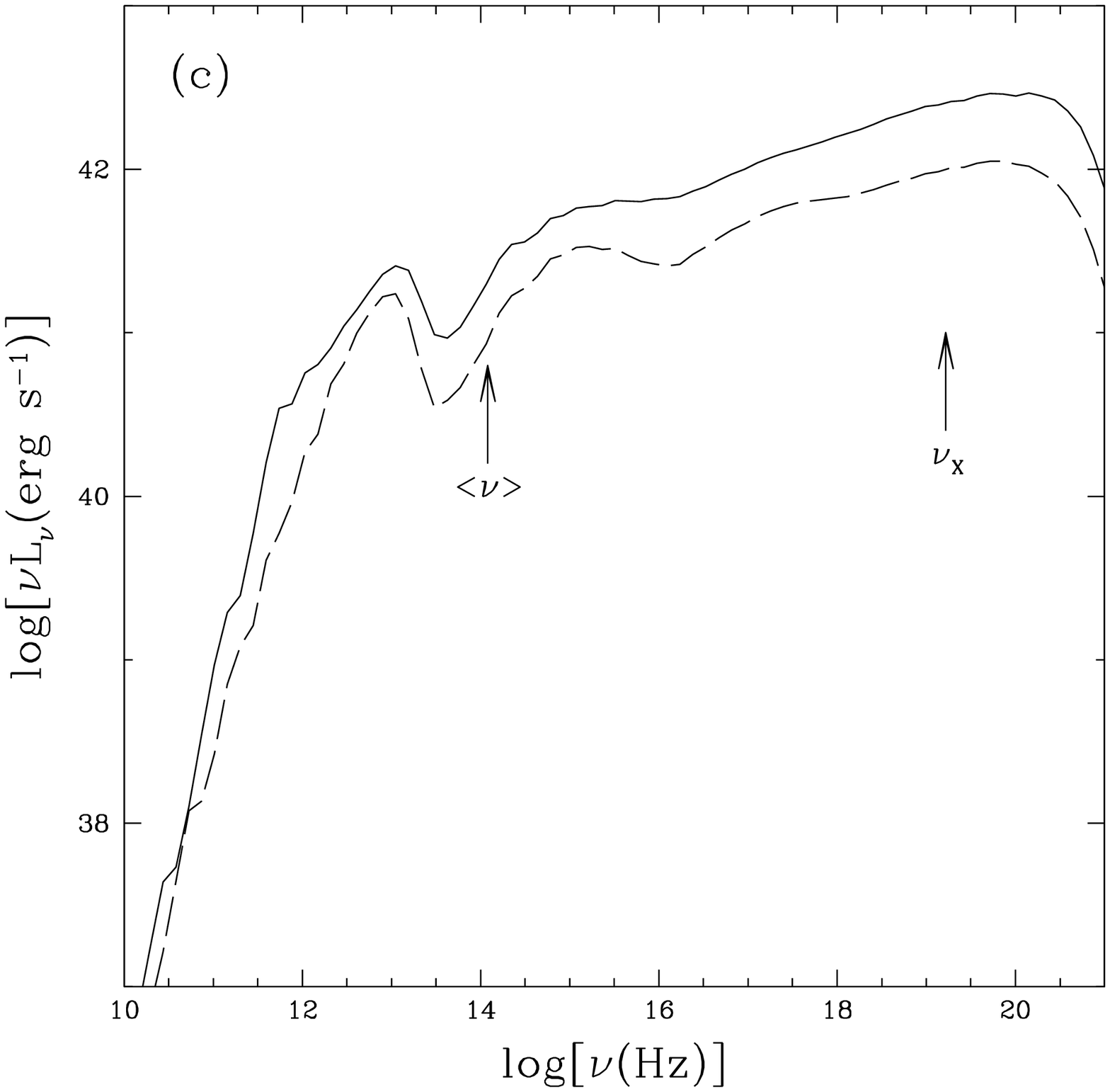}
\caption{Compton effect for a supermassive black hole, with
black hole mass $M=10^8~\msun$ and mass accretion rate
$\dot{M}=0.5\left(\frac{r}{10^4r_s}\right)^{0.3}\dot{M}_{\rm Edd}$.
The solid and dashed lines are for the solutions before and after the global
Compton effect is taken into account. (a) The electron temperature
profile. (b) The ratio of the rate of Compton heating/cooling and the turbulent
heating of electrons. (c) The spectrum of the accretion flows.
The two arrows show the values of the average energy and energy-weighted
energy of photons emitted by the self-consistent solution.}
\end{figure*}

We can see from the left panel of Fig. 1 that the Compton effect is
important when $\dot{M}_0\ga 0.1 \dot{M}_{\rm Edd}$. In this case,
its cooling effect can not be neglected. The corresponding accretion
rate at the black hole horizon is $\sim 10^{-2}\dot{M}_{\rm Edd}$
and the corresponding luminosity is $\sim 5\times 10^{-4}L_{\rm
Edd}$. The ``lowest'' value of $\dot{M}_0$ above which Compton
heating effect is important is a function of $r_{\rm out}$. For
$r_{\rm out}=10^4 r_{\rm s}$, this value is $\sim 2\dot{M}_{\rm
Edd}$ and the corresponding luminosity is $\sim 2 \times 10^{-2}
L_{\rm Edd}$. When $r_{\rm out}$ is larger, the value of critical
$\dot{M}_0$ is lower. In reality $r_{\rm out}$ usually has a largest
feasible value. If the ADAF starts out from a transition from a
standard thin disk, $r_{\rm out}$ equals to the transition radius.
If on the other hand the accretion flow starts out as an ADAF such
as in our Galactic center, $r_{\rm out}$ should be determined by the
Bondi radius. Outside of Bondi radius, the effect of Compton heating
is not so clear, because matching an ADAF solution to one with
proper boundary condition at infinity is an unsolved problem.

Our result that Compton scattering heats electrons at large radii while cools
electrons at small radii is qualitatively consistent with both Esin (1997) and
Park \& Ostriker (2001; 2007). However, Esin (1997) found that Compton heating
rate is always even smaller than the Coulomb collision heating rate therefore
is negligible. This is different from our results and Park \& Ostriker
(2001; 2007). The reason may comes from some over-simplifications
and the different (old) ADAF model adopted in Esin (1997).
Overall, we see that for extended solutions ($r_{\rm out}>10^4r_{\rm s}$),
both Compton cooling in the inner parts and Compton heating in the
outer parts dramatically alter the solutions when $L\ga 10^{-2}L_{\rm Edd}$.

\section{The self-consistent solutions}

The above result indicates that we should take into account the
effect of global Compton heating/cooling when we calculate the
global solution of the hot accretion flow when $\dot{M}$ is
relatively large. This has not been studied in Esin (1997) and Park
\& Ostriker (1999; 2001; 2007). We use an iteration method to
achieve this. We first solve the global solution without considering
the global Compton effect, calculating the rate of Compton
heating/cooling at each radius as described above, $q_{\rm comp}$.
We then include this term in the energy equations of electrons: \be
\rho v \left(\frac{d \varepsilon_e}{dr}- {p_e \over \rho^2} \frac{d
\rho}{dr}\right) =\delta q^+ + q_{\rm ie}-q^- + q_{\rm comp}, \ee
and calculate the ``new'' global solution of the accretion flow
based on this ``new'' equation. Then we  get a new Compton
heating/cooling rate. If the new rate is not equal to the guessed
value we replace the guessed value with the new one and repeat this
procedure until they are equal. However, we must emphasize that the
solution obtained by the above approach is actually not exactly
``self-consistent''. We use eq. (13) to calculate $q_{\rm comp}$.
But in eq. (13), only the local Compton scattering is considered
when calculating $F_{\rm \nu}^{\rm in}+F_{\rm \nu}^{\rm out}$, and
the global scattering is difficult to include because we don't know
the spectrum emitted at other radii which again requires to consider
global scattering. This is difficult to deal with even using the
iteration approach. The best way to solve this problem is by Monte
Carlo simulation combined with iteration method. This is beyond the
scape of this paper and will be our next work. On the other hand, we
believe our result should be a good zeroth-order approximation to the
real solution.

Bearing this in mind, Figs. 2\&3 show the calculation results. Figs. 2 (a--c) are
 for a stellar mass black hole with black hole mass
$M=10~\msun$ and $\dot{M}=(r/50~r_{\rm s})^{0.3}
\dot{M}_{\rm Edd}$. Fig. 2(a) shows the electron temperature of the
global solution without (solid line) and with (dashed line) the global
Compton scattering effect included. Because Compton scattering plays
a cooling role at small radii, we see that the electron temperature decreases after
the Compton effect is taken into account as we expect.

Fig. 2(b) shows the ratio of Compton heating and local viscous
heating of the electrons, $q_{\rm comp}/q_{\rm vis,e}$, before
(solid line) and after (dashed line) the global Compton effect is
included. It is interesting to note that at $r\ga 4r_s$, the
absolute value of this ratio is between $1-4$. Since we typically
have $q_{\rm ie} \ll q_{\rm vis,e}$, this implies that the
right-hand-side of eq. (15) is negative in that region, i.e., the
viscous heating of electrons is smaller than its radiative cooling
($q^--q_{\rm comp})$. In another words, the energy advection of
electrons plays a heating role, just like the ions in the LHAF
solution (Yuan 2001). In the inner region of $r\la 4r_s$, where most
of the radiation comes from, the absolute value of $q_{\rm
comp}/q_{\rm vis,e}\sim 0.5$. We find in this case $(-q_{\rm
comp})\sim q^-$. So in the innermost region the viscous heating of
electrons is equal to its radiative cooling ($q^--q_{\rm comp})$.

Obviously, after taking into account the global Compton cooling, the
radiative efficiency will increase for a given accretion rate. But
this does not mean that the highest luminosity $L_{\rm max}$ a hot
accretion flow can produce will increase. The main heating mechanism
of electrons are viscous heating and compression work (the second
term in the right-hand side of eq. 15) while the main cooling comes
from ($q^--q_{\rm comp}$). The highest accretion rate beyond which a
hot solution no longer exists is determined by the balance between
heating and cooling. The heating term is roughly proportional to
$\dot{M}$ while the cooling term roughly to $\dot{M}^2$ since
Compton scattering is a two-body collision process. This is why a
hot accretion solution has a highest $\dot{M}$. Obviously, when the
global Compton cooling $q_{\rm comp}$ is included, the cooling
becomes stronger compared to the case of only local cooling $q^-$,
thus the balance between cooling and heating will occur at a lower
$\dot{M}$. Actually $\dot{M}_0=L_{\rm Edd}/c^2$ as shown in Fig. 2
is almost the highest accretion rate at which we can get the
self-consistent hot solution, which is a factor of 2-3 lower than
the highest rate when the global Compton effect is not taken into
account. When $\dot{M}_0$ is higher, we find that we are not able to
get the self-consistent solution since the flow will collapse due to
the strong radiative cooling. The decrease of the highest $\dot{M}$
results in the decrease of $L_{\rm max}$. Our calculation shows that
$L_{\rm max}$ decreases by a factor of $\sim $ 2, i.e., we now have
$L_{\rm max}\sim 3\%L_{\rm Edd}$. 

We now check how different the spectrum produced by the
self-consistent solution is compared to the spectrum produced by the
``old'' solution. Fig. 2(c) shows the spectra from the hot accretion
flow before (solid line) and after (dashed line) the Compton effect
is taken into account. We see that both the luminosity and the
cutoff energy of the spectrum (i.e., the corresponding frequency of
the peak of the spectrum) decrease because of the global Compton
cooling. This is of course because the electron temperature of the
self-consistent solution decreases compared to the ``old'' solution.
We would like to emphasize again that only local seed photons are
considered when we calculate the spectrum although we do consider
the global scattering in calculating the dynamics. Our calculation
shows that $-q_{\rm comp} \sim q^-$, so we expect that when the
global Compton scattering is considered the luminosity of the
``exact'' self-consistent solution will be $\sim 2$ times higher
than that shown by the dashed line in Fig. 2(c). But the slope and
the cutoff energy will not change because they are irrelevant to the
amount of seed photons.

Figs. 3 (a--c) are similar to Figs. 2 (a--c), but are for a supermassive black hole
with $M=10^8~\msun$ and accretion rate
$\dot{M}=0.5(r/10^4r_s)^{0.3}\dot{M}_{\rm Edd}$.
We see from the figures that the electron temperature decreases
after the global Compton scattering effect is taken into account as
we expect, because in most region the Compton scattering will cool
the electrons. Correspondingly, the luminosity of the accretion flow
also decreases by roughly a factor of 2 and the cutoff energy of the
spectrum also becomes smaller.

We have also calculated the average energy of the photons emitted by
the self-consistent solution shown in Fig. 3 (c), $h<\nu> ~\equiv
\int L_{\nu}d\nu/\int (L_{\nu}/h\nu)d\nu$, and the corresponding
energy of the radiation temperature $\theta_{\rm x}$ at $r_{\rm
out}$, $h\nu_{\rm x} \equiv m_ec^2\theta_{\rm x}$. The results are
$\sim$ 1~eV and 100~keV respectively, and they are shown by two
arrows in Fig. 3 (c). These values are much higher than that of a
typical quasar spectrum where, e.g., $h\nu_{\rm x}$ is only several
keV (Mathews \& Ferland 1987; Sazonov et al. 2004).

The spectrum shown in Fig. 3 (c) extends to very high energy, $\ga$ MeV.
Observationally, the $e$-folding energy of the average power-law X-ray spectrum
observed by {\it Ginga}, OSSE, and {\it EXSOSAT} of radio-quiet Seyfert 1s
is $E_c=0.7^{+2.0}_{-0.3}$ MeV (Zdziarski et al. 1995; Gondek et al. 1996),
which is consistent with the model given the (large) error bar.
Better data is required to constrain the theoretical model.


As we state in \S2.2, Compton heating effect at large radii is
another obstacle for us to obtain the self-consistent solution. For
$\dot{M}_0=\dot{M}_{\rm Edd}$, if $r_{\rm out} \ga 10^5r_s$, we find
that the Compton heating effect around $r_{\rm out}$ is so strong
that the equilibrium temperature of electrons would be higher than
the virial value defined as $5/2~k~T_{\rm vir} = GMm_p/r$, which
will in turn make the ion temperature also higher than the virial
one due to the Coulomb coupling between them. In this case, the gas
is unbound thus can not be accreted. The corresponding highest
luminosity in this case is $\sim 2 \% L_{\rm Edd}$.
This value is similar to that obtained by Ostriker et al. (1976) and
Park \& Ostriker (2001). From Figs. 1 \& 3(b), we expect that
$\dot{M}_0$ and the critical $r_{\rm out}$ (signed as $r_{\rm
virial}$) beyond which the equilibrium temperature is higher than
the virial temperature are roughly anti-correlated, i.e., a lower
$\dot{M}_0$ corresponds to a larger $r_{\rm virial}$. Exact
estimation of the relation between $\dot{M}_0$ and $r_{\rm virial}$
is not straightforward. This is because we need to know the
radiation temperature $\theta_{\rm x}$ as a function of $\dot{M}_0$
which requires numerical calculations. Note that from Figs. 1 \&
3(b) the minimum value of $r_{\rm virial}$ should be larger than
$\sim 5 \times 10^3 r_{\rm s}$.

Although no steady self-consistent solution exists due to the strong Compton
heating at and beyond $r_{\rm virial}$, an ``oscillation'' of
the activity of the black hole
is expected (e.g., Cowie, Ostriker, \& Stark 1978;
Ciotti \& Ostriker 2007). When the accretion rate is high,
only the gas inside of $r_{\rm virial}$ can be accreted. This active phase will
last for a timescale of the accretion timescale at $r_{\rm virial}$.
Then all the gas will be used up and the active phase stops. In this case,
Compton heating also stops so the gas outside of
$r_{\rm virial}$ will be cooled by radiation and
be accreted again and the cycle repeats.
The time the non-active phase will last is determined by the
radiative timescale of the gas at $r_{\rm virial}$,
since it is longer than the accretion timescale there. An alternative
consequence of the strong Compton heating at large radii is that
the accretion can be self-regulated by irradiating the outer flow
(Shakura \& Sunyaev 1973). That is, the strong Compton heating
will not completely stop the accretion, but only decrease the accretion rate.
This then reduces the energy release in the inner part, which in turn
reduces the irradiation. A multi-dimensional numerical simulation is 
required to solve this issue and accurate time-dependence is needed
as well since steady solutions may not be stable. 

For the massive black holes seen in the nuclei of most
galaxies the Compton heated interruption of the high luminosity
states should be typical if a hot accretion flow exists there.
We now know that the accretion flow in low-luminosity AGNs
is of this type (see Yuan 2007 and Ho 2008 for reviews). For luminous
AGNs such as quasars, although people incline to think it is 
a standard thin disk which is optically thick, many problems remain 
for this model (e.g., Shlosman, Begelman, \& Frank 1990; Koratkar 
\& Blaes 1999). If the actual accretion flow is radially optically 
thin to Compton scattering, similar to the hot accretion flow, 
our analysis applies.

This kind of oscillation does not apply
to stellar mass black holes in our Galaxy. This is because the prerequisite
for such oscillation is that the accretion rate is large and the hot
accretion flow extends to large radii. For a stellar mass black hole, the
accretion material comes from the companion star and it starts out
as a standard thin disk. In the hard state the standard disk does
not extend to the innermost stable circular orbit but is replaced by
a hot accretion flow within a transition radius $r_{\rm tr}$. However,
when the accretion rate is high, $r_{\rm tr}$ is small (Yuan \& Narayan 2004).
So Compton scattering cools rather than heats the hot accretion flow.  But on
the other hand, the photons emitted by the hot accretion flow will heat
the cool electrons in the standard disk at $r_{\rm tr}$
and thus will change the dynamics of the transition.
This effect has never been noted and could be a topic of future work.

\section{Summary and Discussion}

For a geometrically thick and optically thin hot accretion flow,
the photons can travel for a long distance without being
absorbed, and thus be able to heat or cool electrons via Compton
scattering. We investigate this global Compton scattering
effect and find that for an accretion rate described by
$\dot{M}=\dot{M}_0(r/r_{\rm out})^{0.3}$ the Compton cooling effect will
be important when $\dot{M}_0 \ga 0.1L_{\rm Edd}/c^2$;
while the Compton heating effect will be important when
$\dot{M}_0 \ga 2L_{\rm Edd}/c^2$ and $r_{\rm out}=10^4r_{\rm s}$.
Specifically, the scattering heats electrons at $r>5 \times 
10^3 r_{\rm s}$ while cools electrons at $r<5 \times 10^3 r_{\rm s}$. 
If $r_{\rm out}$ is larger, the critical $\dot{M}_0$ above which 
the Compton heating effect is important will become lower.

We have successfully obtained the self-consistent steady solution with
this effect included for $\dot{M}_0 \la L_{\rm Edd}/c^2$ 
and $r_{\rm out}=50r_s$. But when $\dot{M}_0\ga L_{\rm Edd}/c^2$ 
and $L>2\%L_{\rm Edd}$  we fail because of the strong
radiative cooling (local plus global Compton scattering).
It is also difficult to get the self-consistent solution when
$\dot{M}_0\ga L_{\rm Edd}/c^2$ ($L>1\%L_{\rm Edd}$)
 and $r_{\rm out}\ga 10^5 r_s$. This is because
in this case the Compton heating is so strong at and beyond $r_{\rm out}$
that the equilibrium electron temperature there will be higher
than the virial temperature. More generally we expect that the radius where
the equilibrium temperature due to the Compton heating is equal
to the virial temperature, $r_{\rm virial}$, is anti-correlated with
$\dot{M}_{\rm out}$. We argue that
the black hole will manifest an oscillation of the activity in the case that
we fail to get the steady solution.  The period
will be the radiative timescale of the gas at $r_{\rm virial}$.

All our discussions so far are for a one-dimensional (but not spherical)
accretion flow. Although
big uncertainties exist for the vertical structure of accretion flow,
we are certain that when $\dot{M}$ is high, the scattering will be important,
and consequently much of the luminosity will ``leak out'' perpendicular
to the accretion flow as in the standard thin disk. This will have two
effects. One is that the highest luminosity up to which we can get the
self-consistent solution with the global Compton effect included will be higher.
In addition, the Compton heating will be stronger in the vertical direction
than in the equatorial plane of the flow. As a result strong wind will 
be launched as pointed out by Park \& Ostriker (2001; 2007) and 
found by Proga et al. (2008). All of the described effects are likely 
to become significant for AGN accretion flows having $L>10^{-2}L_{\rm Edd}$
and optically thin (ref. \S 3 for discussion of this possibility in
luminous AGNs), which, we know from recent applications of the 
Soltan argument (Yu \& Tremaine 2002), are the phases during which most 
massive black hole growth occurs.

\acknowledgements

We thank Ramesh Narayan, Myeong-Gu Park, Eliot Quataert, and the
referee, Andrzej Zdziarski, for their useful comments on the work.
This work was supported in part by the Natural Science Foundation of China
(grants 10773024, 10833002, and 10825314), One-Hundred-Talent Program
of Chinese Academy of Sciences,
and the National Basic Research Program of China (grant 2009CB824800).

{}

\clearpage

\end{document}